\documentstyle[preprint,aps,epsf,epsfig,floats]{revtex}
\begin{document}
\draft \preprint{\hbox{UM PP\#03-009}}
\title{Does A Narrow Tetraquark $c\,c\,\bar{u}\,\bar{d}$ State Exist?}
\author{Boris A. Gelman$^{*}$ and Shmuel Nussinov$^{**}$}
\address{$^{*}$ Department of Physics, University of Maryland, College Park\\
College Park, MD 20742-4111, USA and \\
Department of Physics, University of Arizona, Tucson, AZ 85721, USA \\
$^{**}$ Sackler Faculty of Science, Tel Aviv University, Tel Aviv,
Israel}
\maketitle
\begin{abstract}
The existence of a shallow or virtual tetraquark state, 
$c\,c\,\bar{u}\,\bar{d}$, is discussed. Using the putative masses for the 
doubly charmed baryons ($c\,c\,u/c\,c\,d$) from SELEX, the mass of the 
$c\,c\,\bar{u}\,\bar{d}$ state is estimated to be about $3.9 \, GeV$, only
slightly above the $D\, D^*$ threshold. The experimental signatures for
various $c\,c\,\bar{u}\,\bar{d}$ masses are also discussed.
\end{abstract}

\def\N{N_c}
\def\m{m_Q}
\def\L{\Lambda_{QCD}}
\def\bL{\bar{\Lambda}}
\def\l{\lambda}
\def\c{\cite}
\def\r{\ref}
\def\qb{\bar{q}}
\def\qbp{\bar{q'}}
\def\e{\epsilon}
\def\tq{c\,c\,\bar{u}\,\bar{d}}
\def\cud{c\,u\,d}
\def\sud{s\,u\,d}
\def\cubar{c\,\bar{u}}
\def\subar{s\,\bar{u}}
\def\ccu{c\,c\,u}
\def\ccd{c\,c\,d}
\def\tetra{Q\,Q'\,\bar{q}\, \bar{q}'}

\section{Introduction \label{intro}}

Multiquark exotic hadrons different from the ordinary mesons or
baryons have been discussed and searched for many years 
\cite{jaffe1,jaffe2,Lipkin1,Lipkin2,WI1,WI2,richard1,richard2}.
Motivated by the possible discovery of doubly charmed
$c\,c\,u/c\,c\,d$ baryons, \cite{SELEX}, we reconsider the
$c\,c\,\,\bar{u}\,\bar{d}$ tetraquark---the doubly charmed exotic
state. We find that such a $J^P=1^{+}$ state is likely to exist
below or near the $D^{*}\, D$ thresholds, and may be the first
stable/narrow exotic state to be discovered.\footnote{Much effort was devoted 
in the past to search for hexaquark and pentaquark states 
\cite{Lipkin1,OZ,nuss,richard3}.}

\section{$\tetra$ States in the Heavy Quark Limit
\label{hql}}

\begin{figure}[ht]
\begin{center}
\epsfig{figure=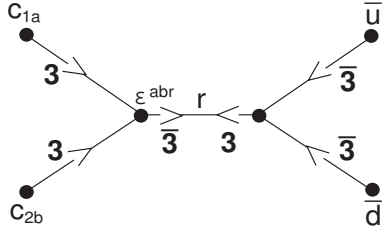,height=3cm,width=5cm,angle=0}
\bigskip
\caption{The color coupling pattern adopted here. Two charm quarks
$c_1\,,c_2$ couple antisymmetricly to an intermediate ${\bf
\bar{3}}$ and $\bar{u}\,,\bar{d}$ couple to ${\bf 3}$.}
\label{fig1}
\end{center}
\end{figure}

Let $T(Q\,Q'\,\bar{q}\,\bar{q}')$ denote a putative tetraquark
state (it is denoted as $X_c$ in \cite{Lipkin2}) consisting of two heavy
quarks---$Q,\,Q'= c$ or $b$---and
two light quarks---$q,\,q'= u,\,d$. We are interested in genuine
four quark ``one bag states'' with a ``connected color network''.
Figure~\ref{fig1} indicates one such state where
$Q_{\alpha}\,Q'_{\beta}$ are combined via
$\epsilon_{\alpha\beta\gamma}\, Q_{\alpha}\,Q'_{\beta}$ into
$(\bf \bar{3})_\gamma$ color diquark, and $q_{\mu}\,q'_{\nu}$ make an
$\epsilon_{\mu\nu\gamma}\, \bar{q}_{\mu}\,\bar{q}'_{\nu}=(\bf 3)_\gamma$---an
anti-diquark. The ${\bf \bar{3}}$ and ${\bf 3}$ combine to give
an overall color singlet state. Note that according to the well-known
arguments, \cite{LN1,LN2}, in the large $N_c$ limit such a four-quark state is
unstable against a decay into the two mesons, $D$ and $D^*$ for instance.
Alternatively, however, one can assume that the state $T(\tq)$ in
the large $N_c$ limit corresponds to the state containing $N_c -1 $ heavy
quarks, which combine into ${\bf \bar{N}_c}$ representation of $SU(N_c)$
color, and $N_c -1$ light quarks combined into ${\bf N_c}$. According to the
standard large $N_c$ counting rules the binding energy of such a
$(2 N_c-2)$-exotic state, \cite{chow,Cohen}, is of order
${\cal O}(N_{c}^{0})$,  {\it i.e.} of the same order as the meson masses. 

The identity $\epsilon_{\alpha\beta\gamma}\,\epsilon_{\sigma\tau\gamma}=
\delta_{\alpha\sigma}\,\delta_{\beta\tau} \, - \,
\delta_{\alpha\tau}\,\delta_{\beta\sigma}$ allows one to express the
above tetraquark state as a superposition of two meson
states, $(Q\,\bar{q})$ etc., which are separately color singlets.
\begin{equation}
T\{Q\,Q' \to {\bf \bar{3}},\, \bar{q}\,\bar{q}'\to {\bf 3}\} =
|\,(Q\,\bar{q})_1\, (Q'\,\bar{q}')_1 \,\rangle -
|\,(Q\,\bar{q}')_1\, (Q'\,\bar{q})_1 \,\rangle\, .
\label{identity}
\end{equation}
If the $Q\,Q'$ (and $\qb\,\qb'$) colors are coupled symmetrically
to ${\bf \bar{6}}$ (and ${\bf 6}$) the analog of the
Eq.~(\ref{identity}) will have a plus relative sign.

\begin{figure}[t]
\begin{center}
\epsfig{figure=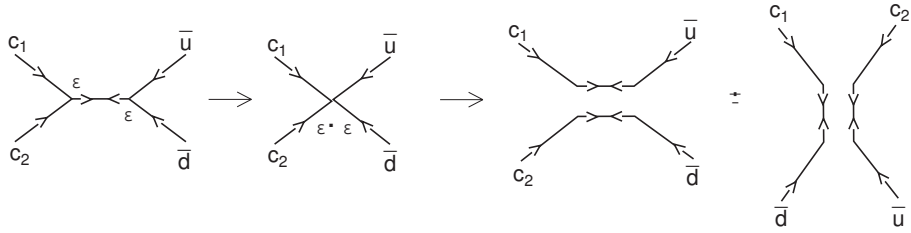,height=3cm,width=12cm,angle=0}
\bigskip
\caption{The color flux vortices and their evolution as the
tetraquark state separates into two $D + D^*$ states.}
\label{fig2}
\end{center}
\end{figure}

In a ``string picture'' the chromoelectric fluxes are squeezed
into thin ``vortices'' connecting the various (anti)quarks and/or
junction points. In this case the various lines in Fig.~\ref{fig1}
describe not only the color coupling but also the actual layout
of the strings. The transition between the tetraquark (color
connected---``one-bag'' state) to the two-meson state can be
pictorially described by shrinking the string bit connecting the
two junctions, and then annihilating them via the above $\e \cdot \e$
contraction (Fig.~\ref{fig2}). This naively would suggest the
two-meson state, in which the above string bit has been
eliminated, is lighter than the tetraquark state rendering the
latter unstable since $T(Q\,Q'\,\bar{q}\,\bar{q}') \to Q\,\qb +
Q'\,\qb'$ would be kinematically allowed.

However, the naive string picture may not apply to the ground
state hadron considered here. Indeed we can directly show that
$T(Q\,Q'\,\qb\,\qb')$ {\it is stable} in the heavy quark limit 
($m_{Q,Q'} \to \infty$). The $Q$ and $Q'$ would then bind into a 
${\bf \bar 3}$ via
 the perturbative one gluon exchange. The essentially 
coulombic interaction yields a binding energy ${\cal
O}(\alpha_{s}^{2})\,\m/2$ (for $\m=m_{Q'}$). Once $\m$ is
sufficiently large this binding exceeds hadronic energy scales
and possible bindings in the heavy-light $Q\,\qb$ mesonic
systems. This then ensures the stability against decay into two
such mesons.

Unfortunately for this mechanism to generate stable tetraquarks
$\m$ (much) larger than ($m_c$) $m_b$ is required. Detailed, and to
some extent model dependent, considerations are thus required to
motivate a $c\,c\,\bar{u}\,\bar{d}$ state near or below $D^* \,D$
threshold. An alternative approach, which can directly address the
stability of the physical $1^{+}$ ($c\,c\,\bar{u}\,\bar{d}$)
state against a decay into $D\, D^*$ or $D^* \, D^*$, starts with
these charmed mesons and attempts to form bound states via the
potential due to light mesons, particularly one-pion exchange.
Following T$\ddot{\rm{o}}$rnqvist, \cite{Tdeuson} we will call
such ``deutron like'' or
``molecular'' bound states of two mesons, ``deusons''. The
ranges and strengths of such potentials are independent of $\m$
since $D \, D^* \,\pi$ or $D^* \, D^*\,\pi$ couplings, for instance,
depend only on the light quark degrees of freedom. Thus, if the
two heavy mesons are attracted by these potentials, binding is
again guaranteed in the heavy quark limit as the kinetic energy
and centrifugal barriers for $\ell \ne 0$ waves vanish like
$1/\m$.

\section{Do $D^* \, D$ bound deusons exist? \label{DD}}

Previous calculations utilizing OPEP (one-pion exchange potential)
disagree on this issue. Thus T$\ddot{\rm{o}}$rnqvist 
\cite{Tdeuson,Tdeuson2,T} finds
$B^* \,B$, $D^* \, D$ and even $K^* \, K$ bound states whereas Manohar
and Wise \cite{MW} find only $B^* \, B$ bound states. The
difficulty stems from the fact that much of the binding is due to
the short range part of the potential.\footnote{The central part
of the OPEP in the $D^{(*)} \, D^{(*)}$ channel has the form:
$${(\vec{\hat{\e}}_1 \cdot \vec{q}) \, (\vec{\hat{\e}}_2 \cdot \vec{q})
(\vec{\tau}_1 \cdot \vec{\tau}_2) \over q^2 + \mu^2} \, $$ with
$\mu^2 = m_{\pi}^{2} - (m_{D^*} - m_D)^2$ in $D^* - D$ and $\mu^2
= m_{\pi}^{2}$ in $D^* - D^*$. Averaging over the polarizations
we get an expression like $q^2/(q^2+\mu^2) \approx 1-
\mu^2/(q^2+\mu^2)$. The second piece becomes negligible as $\mu
\to 0$ (except at $q=0$) and the first piece contributes in
configuration space a $\delta$-function which clearly cannot be
utilized in a reliable way for binding an extended object of
interest.} The derivative pion coupling generates in the tensor
part of the OPEP a $e^{-\mu r}/r^3$ term which is singular at
short distance. Cutting off the OPEP at some distance $r_0$ is therefore
required. The fact that Manohar and Wise (conservatively (?))
chose $r_0 \approx 1/2m_{\pi}$ and T$\ddot{\rm{o}}$rnqvist (boldly (?))
takes
$r_0 \approx 1/4m_{\pi}$ is the likely reason for their
disagreement on bound $D^* \, D$ deusons.

We do not believe that this issue can be settled. Yet the fact
that the binding of this system by OPEP alone is not guaranteed
may make the problem even more interesting. To decide the issue
of a physical bound state with $\tq$ flavor we may need the
genuine four quark---``one bag'' component of the state---where
nontrivial aspects of QCD are operative. Indeed, once the
distance between the $D$ and $D^*$ mesons defined, say, by the distance
between the respective charmed quarks, is smaller than the size of
$D$ or $D^*$ we can no longer treat the system as two separate
hadrons exchanging light mesons. Rather, we need to revert to the
``one bag'' tetraquark description. The various quark-(anti)quark
interactions in this state may then supply just the extra
attractive interaction at (relatively) short distances needed in
order to bind the system.

The same physical $\tq$ state would then be a $T(\tq)$ at short
distances, $r \leq r_0 (\approx 0.7 fm)$ and a $D^*\,D$ deuson
for $r\geq r_0$. Neither deusons nor tetraquarks are true ground
states. The basic variational principle implies that mixing will
{\it lower} the true ground state energy below the lowest
energies found in each sector separately.

\section{The $\tq$ tetraquark \label{ccud}}

Lacking a consistent first principle computational
framework we appeal to the vast existing lore and literature.
Thus to approach the problem of ``color connected single bag'' states 
one utilizes:\\
 i) ``constituent'' massive quarks $m_u \approx m_d
\approx 350 \, MeV$, $m_c \approx 1.6\, GeV$ ;\\
 ii) appropriate $q/\bar{q}-q$ long range interactions, or alternatively,
 an overall bag which confines the quarks into a single
 state; \\
 iii) the chromomagnetic hyperfine pairwise interactions:
\begin{equation}
H.F. \equiv {\cal H}_{ij} \approx -{1\over m_i \, m_j} \,
``{|\Psi_{ij}(0)|^{2}}'' (\vec{\sigma}_i \cdot \vec{\sigma}_j) \,
(\vec{\l}_i \cdot \vec{\l}_j) \,, \label{hf}
\end{equation}
where $\sigma_i$/$\l_i$ are the spin/color matrices of the
(anti)quarks, and ``${|\Psi_{ij}(0)|^{2}}$'' is the relative wave
function at zero separation for a $q_i \bar{q}_j$ meson and more
generally the probability of overlap of the two (anti)quarks
considered.

Rather than attempting an {\it ab initio} calculation we adopt a more
phenomenological approach. It utilizes known masses and insight
from successful past calculations instead based on (i)-(iii) above
in order to extrapolate to the $\tq$ mass.

We focus on the $\tq$ $I=0$ state (rather then
$c\,c\,\bar{u}\,\bar{u}$ ($c\,c\,\bar{d}\,\bar{d}$) $I=1$ states)
since in both the tetraquark and deuson approach it is more
strongly bound. The $H.F.$ hyperfine interaction, Eq.~(\ref{hf}), strongly
favors ${^1}S$ $\bar{u}\,\bar{d}$ which is a ${\bf 3}$ of color. The
resulting anti-symmetry in color and separately in spin does then
require flavor antisymmetry, {\it i.e.} isosinglet $\bar{u}\,\bar{d}$
state, in order to maintain overall Fermi statistics. Note that the
$c_1\,c_2$ quarks, which are in ${\bf \bar{3}}$ of color, must then form a
spin triplet. This suggests an overall $s$-wave $1^{+}$ singlet $\tq$ state
whose lightest putative decay channel is indeed $D\,D^*$. Also in the deuson
approach the attractive OPEP is three times as strong
in the $I=0$ than in the $I=1$ channel.

We thus consider the following putative double difference relation,
\begin{equation}
\left (m_{(\tq){1^+}}\,-
\,m_{(c\,c\,u){1/2}^{+}}\right)\,-\,\left(m_{
\Lambda_c}\,-\,m_{D^{0}}\right) =
0\,.\label{main}
\end{equation}
It assumes that the extra energy required for replacing a $u$ (or
$\bar{u}$) quark by a $u\,d$ ($\bar{u}\,\bar{d}$) diquark in the
presence of a $c$ quark or $(c\,c)_{\bf \bar{3}}$ diquark is the
same. It is clearly true in the heavy quark limit. Indeed
in this limit the compact tightly bound $Q\,Q$ pair is like
a heavier antiquark flavor. Equation~(\ref{main}) would then be 
analogous to the relation inspired by the heavy quark symmetry: 
\begin{equation}
\left (m_{\Lambda_b}\,-
\,m_{B}\right)\,-\,\left(m_{\Lambda_c}\,-\,m_{D}\right) =
0\,,\label{HQSYM}
\end{equation}
which holds very well.

To further motivate Eq.~(\ref{main}) we note that quark masses
and almost all pairwise interactions cancel out via the double
difference. The interactions between the $c_1$ and $c_2$ quarks in
the $c\,c = {\bf \bar{3}}$ diquark in $\tq$ and in $c\,c\,u$ match
as well as the interactions between the $\bar{u}\,\bar{d} = {\bf
3}$ ($1s$ of spin) in $\tq$ and the similar $u\,d$ pair in
$c\,u\,d$.

The $H.F.$ interaction between the $c$ quark and the $u\,,d$
quarks in $\Lambda_c =(c\,u\,d)$ cancel out since
$\vec{\sigma}_u = -\,\vec{\sigma}_d$ implies
\begin{eqnarray}
&& \left(\vec{\sigma}_u \cdot \vec{\sigma}_c
\right)\left(\vec{\l}_u \cdot \vec{\l}_c\right) +
\left(\vec{\sigma}_d \cdot \vec{\sigma}_c \right)\left(\vec{\l}_d
\cdot \vec{\l}_c\right) = \left(\vec{\sigma}_u \cdot
\vec{\sigma}_c \right)\left(\left(\vec{\l}_u -\vec{\l}_d \right)
\cdot \vec{\l}_c\right) \nonumber \\
&& = - \left(\vec{\sigma}_u \cdot \vec{\sigma}_c
\right) \left(\vec{\l}_{u}^{2}
-\vec{\l}_{d}^{2} \right) = 0 \,, \label{cancel1}
\end{eqnarray}
where we utilized the color neutrality condition,
$\vec{\l}_u + \vec{\l}_d + \vec{\l}_c =0$, for
$\Lambda_c$.

Similarly the $H.F.$ interactions between the charmed quarks and the
$\bar{u}\,,\bar{d}$ quarks in $T(\tq)$ cancel. Apart from
an overall common factor $|\Psi_{uc}(0)|^2 /m_u\, m_c$ we have,
using again $\vec{\sigma}_u =-\,\vec{\sigma}_d$,
\begin{eqnarray}
&& \left(\vec{\sigma}_{c1} \cdot \vec{\sigma}_{\bar{u}}
\right)\left(\vec{\l}_{c1} \cdot \vec{\l}_{\bar{u}}\right) +
\left(\vec{\sigma}_{c1} \cdot \vec{\sigma}_{\bar{d}}
\right)\left(\vec{\l}_{c1} \cdot \vec{\l}_{\bar{d}}\right)
\nonumber \\
&& + \left(\vec{\sigma}_{c2} \cdot \vec{\sigma}_{\bar{u}}
\right)\left(\vec{\l}_{c2} \cdot \vec{\l}_{\bar{u}}\right) +
\left(\vec{\sigma}_{c2} \cdot \vec{\sigma}_{\bar{d}}
\right)\left(\vec{\l}_{c2} \cdot \vec{\l}_{\bar{d}}\right)
\nonumber \\
&& = \left(\vec{\sigma}_{c1} \cdot \vec{\sigma}_{\bar{u}}
\right)\left(\vec{\l}_{\bar{u}} -\vec{\l}_{\bar{d}} \right) \cdot
\vec{\l}_{c1} +  \left(\vec{\sigma}_{c2} \cdot
\vec{\sigma}_{\bar{u}} \right)\left(\vec{\l}_{\bar{u}}
-\vec{\l}_{\bar{d}} \right) \cdot \vec{\l}_{c2} = \hat{O}_A \,.
\label{cancel2}
\end{eqnarray}
The last expression is an operator $\hat{O}_A$ anti-symmetric
under the exchange of the color degrees of freedom of $\bar{u}$
and $\bar{d}$. In the tetraquark state the $\bar{u}$ and $\bar{d}$
colors are also coupled antisymmetrically  in forming the ${\bf
3}$ (anti)diquark. As a result the total $H.F.$ interaction energy between
the light and the heavy quarks in the tetraquark has the
form $\langle \, T_A \,|\hat{O}_A| \, T_A\,\rangle$ which vanishes
since upon an exchange of color indices of $\bar{u}$ and $\bar{d}$
this matrix element changes sign.

To complete motivating Eq.~(\ref{main}) we still need to show
that the $c\,u$ (and $c\,\bar{u}$) $H.F.$ interactions in
$\Xi_{ccu} ({1/2}^{+})$ and in $D^0(0^-)$ match. The
latter is:
\begin{equation}
 -{1\over m_u \, m_c} \, |\Psi_{uc}(0)|^{2}
(\vec{\sigma}_c \cdot \vec{\sigma}_u) \, (\vec{\l}_c \cdot
\vec{\l}_u) \approx -{3\over 4} \, \l^2 {|\Psi_{uc}(0)|^{2} \over
m_u \, m_c} \,, \label{hf2}
\end{equation}
where we used $\vec{\sigma}_c + \vec{\sigma}_u = 0$ and
$\vec{\l}_c + \vec{\l}_u = 0$.

The $H.F.$ interactions in the doubly charmed baryon are however:
\begin{eqnarray}
& - & {1\over m_u \, m_c} \, |\Psi_{uc}(0)|^{2}
\left\{(\vec{\sigma}_u \cdot \vec{\sigma}_{c1}) \,
\left(\vec{\l}_u \cdot \vec{\l}_{c1}\right) + (\vec{\sigma}_u
\cdot \vec{\sigma}_{c2}) \, \left(\vec{\l}_u \cdot
\vec{\l}_{c2}\right) \right\}  \nonumber \\
& = & {1\over2}\, {|\Psi_{uc}(0)|^{2}\over m_u \, m_c} \,
\{\left(\vec{\sigma}_u \cdot (\vec{\sigma}_{c1} +
\vec{\sigma}_{c2})\right) \, \left(\vec{\l}_u \cdot \left(
\vec{\l}_{c1} + \vec{\l}_{c2}\right)\right) \nonumber \\
& + & \left(\vec{\sigma}_u \cdot (\vec{\sigma}_{c1} -
\vec{\sigma}_{c2})\right) \, \left(\vec{\l}_u \cdot \left(
\vec{\l}_{c1} - \vec{\l}_{c2}\right)\right) \} \,. \label{hf3}
\end{eqnarray}
The second term above vanishes since, again,
$$\vec{\l}_u \cdot \left(
\vec{\l}_{c1} - \vec{\l}_{c2}\right) \approx \left(\vec{\l}_{c1}
+\vec{\l}_{c2} \right) \cdot \left( \vec{\l}_{c1} -
\vec{\l}_{c2}\right) = \l_{c1}^{2} - \l_{c2}^{2} =0 $$ so that we
obtain:
\begin{equation}
 -{1\over 2} \, {|\Psi_{uc}(0)|^{2} \over m_u \, m_c} \,
(\vec{\sigma}_u \cdot (\vec{\sigma}_{c1} + \vec{\sigma}_{c2}))
\cdot \l^2 = - {1\over 2} \, \l^2 {|\Psi_{uc}(0)|^{2} \over m_u \,
m_c} \,, \label{hf4}
\end{equation}
where in the last step we used $\left( \vec{\sigma}_u +
\vec{\sigma}_{c1} + \vec{\sigma}_{c2} \right)^2 =3/4$ (since the
lightest charmed baryon has spin 1/2) and $\left(\vec{\sigma}_{c1}
+ \vec{\sigma}_{c2} \right)^2 = 2$ (since the two charmed quarks
are in $^3 S$ state).

We find that the hyperfine (attractive) interaction in
$\Xi_{ccu} ({1/2}^{+})$ and in $D^0 (0^-)$ do not
exactly match but are slightly stronger in $D^0$ by
$${1\over 4}\, ``H.F.'' = {1\over 4}\, \left(m_{D^*} -
m_{D}\right)\,.$$ Thus by subtracting the physical {\it lighter}
D mass we are actually causing an imbalance in Eq.~(\ref{main})
and an {\it overestimate} of $m_{T(\tq)}$. A corrected equation
should therefore read:
\begin{equation}
m_{T(\tq)}=m_{\Xi_{ccu}} + m_{\Lambda_c} -
m_{D^0} -{1\over 4}\,\left(m_{D^*} -
m_{D}\right) \,. \label{mT}
\end{equation}
While the masses of $\Lambda_c$, $D^0$ and $D^{0*}$ are well known,
this is definitely not the case for the doubly charmed baryons.
The lowest SELEX peak appears in ${c\,c\,u}^{(++)}$ at $3460 \,
MeV$ \cite{SELEX}. Using this value in Eq.~(\ref{mT}) yields
\begin{equation}
m_{T(\tq)} \approx 3.845 \, GeV \,, \label{prediction1}
\end{equation}
which is about $25\, MeV$ {\it below} the $D^*\,D$ threshold.

Unfortunately the lowest peak in ${c\,c\,d}^{(+)}$ is at $3.52 \,
GeV$ and not degenerate with the ${c\,c\,u}$ peak as it should be
by isospin invariance. We therefore choose the ${c\,c\,d}^{(+)}$
peak (which indeed coincides with a (relatively small)
enhancement in the ${c\,c\,d}^{(+)}$ SELEX data) as representing
the true value of the lowest charmed baryon. This is clearly a
more conservative choice as both SELEX peaks are lower than most
previous theoretical predictions. Using this we obtain,
\begin{equation}
m_{T(\tq)} \approx 3.905 \, GeV \, \label{prediction2}
\end{equation}
some $35\, MeV$ {\it above} the $D^*\,D$ threshold.

Various explicit and implicit assumptions were made in order to
obtain Eq.~(\ref{mT}) making for a theoretical uncertainty in the
predicted value of $m_{T(\tq)}$ beyond the $60 \, MeV$
experimental uncertainty discussed above.

a) We used a common overlap probability ``$|\Psi_{ij}(0)|^2$'' for
all $c\,\bar{q}$ or $c\,q$ pairs which corresponds to a
``universal bag radius''. In reality the latter could change as we
go from two to three to four quark systems. The successful
phenomenology of baryon/meson hyperfine splitting suggests that
this may be a weak effect. Furthermore a systematic change will
largely cancel in the double difference relation in Eq.~(\ref{main}).

b) We have restricted our discussion to diquarks coupled to ${\bf
\bar{3}}$ (or ${\bf 3}$) of color only. However in the tetraquark
system we encounter (for the first time) the possibility of
coupling $c_1 \,c_2$ (and $\bar{u}\,\bar{d}$) to ${\bf 6}$ (${\bf
\bar{6}}$) of color---a coupling, which is clearly not allowed in
a baryon. The restriction to the ${\bf \bar{3}}-{\bf 3}$ pattern
may be well justified by the fact that it has a lower energy than
the ${\bf 6}-{\bf \bar{6}}$. Nonetheless
these channels can in principle mix, and allowing for such
admixture to optimize the binding will lower the mass of the
physical state.

c) The admixture of the ``one bag'' tetraquark state above and the
``two bag'' deuson state (which again occurs first in four-quark
exotic states) will also, by the above mentioned variational
argument, tend to lower the energy.

All the above suggests that if the SELEX peak at $3460 \, MeV$ is
indeed the lightest double charmed baryon, we have a $T(\tq)$
state slightly below or slightly above threshold. Specifically, if
\begin{equation}
\e = m_{T(\tq)} - m_{D^*} - m_{D}  \,, \label{e}
\end{equation}
we expect
$$ |\e| \leq 30 \div 60 \, MeV\, .$$

\section{Production and Decays of $T(\tq)$ \label{rate}}

We would next like to argue that if $T(\tq)$ has the above mass it
may well be the first narrow exotic hadron to be discovered. The potential
discovery depends jointly on\\
(i) the rate of $\tq$ production \cite{Moinester}, and \\
(ii) the existence of decay modes which can provide a unique
signature.

The production rates (at hadronic or $e^+ \, e^-$
colliders) of the state $\tq$ of interest are very small as all
the following conditions should be met: \\
(a) Two pairs of charmed quarks $(\bar{c}_1\,c_1)$ and
$(\bar{c}_2\,c_2)$ need to be produced. \\
(b) These pairs should be close spatially. Also $c_1$ from
the first pair, and $c_2$ from the second should have small
relative momenta in order for a
a $c_1\,c_2$ diquark to form. \\
(d) Finally the $c_1\,c_2$ diquark should pick up a
$\bar{u}\,\bar{d}$ (anti)diquark to form $\tq$.

The first two factors (a) and (b) also suppress the production
rate of doubly charmed baryons $c\,c\,u$/$c\,c\,d$. The only
further suppression of the $\tq$ production rate is due to the
need to pick up a $\bar{u}\,\bar{d}$ diquark instead of merely
just one $u$ (or $d$) in the case of $c\,c\,u/c\,c\,d$. This suggests the
following ``double ratio'' relation:
\begin{equation}
\left(R_{(\tq)} / R_{(c\,c\,u)}\right) : \left(R_{(\cud)} /
R_{(\cubar)}\right) =1 \,, \label{RR}
\end{equation}
which is analogous to the double difference relation in
Eq.~(\ref{main}).\footnote{This analogy is very natural in a ``statistical
model'', where the
production rate of any hadronic state $X$ is suppressed by a
Boltzmann factor proportional to $\exp{(-m_{X}/T)}$.
Equation~(\ref{RR}) results then as the exponential of
Eq.~(\ref{main}).}

The ratio of the charmed baryon, $\Lambda_c$, production and that
of $D$'s is roughly the same as in the case of strange
baryons/mesons:
\begin{equation}
R_{(\cud)} / R_{(\cubar)} \approx R_{(\sud)} / R_{(\subar)} =
R_{(\Lambda)} / R_{(K)} \approx {1\over 10}\,. \label{RRstrange}
\end{equation}
Hence we expect from Eq.~(\ref{RR})
\begin{equation}
R_{(\tq)}  \approx {1 \over 10} R_{(\ccu)} \, , \label{RRtetra}
\end{equation}
namely that the $T(\tq)$ production rate is about $1/10$ that of
charmed baryons.

If $T(\tq)$ is to be discovered this tiny production rate needs to
be compensated by striking decay signatures. The decay modes
critically depend on $m_{T(\tq)}$ or $\e$, the separation between
$m_{T(\tq)}$ and the $m_{D}+ m_{D^*}$ threshold, defined in 
Eq. (\ref{e}). 

We will next discuss the different $m_{T(\tq)}$  ranges starting
with the most strongly bound case.

(a) $m_{T(\tq)} \leq 2\, m_{D}$ or $\e \leq m_{D} - m_{D^*}
\approx -\, 140 \,MeV$.\\
In this case---which the above discussion suggests to be
unlikely---we can only have two consecutive weak decays. Strictly
speaking only the second vertex involves an on-shell
reconstructible $D^+$ or $D^0$. However, up to small binding
effects, $2\, m_{D} - m_{T(\tq)}$, the tracks emerging from the
first decay may well correspond to another $D$. 

There is some probability that the charm quark surviving after the
first decay may be inside a $D^*$ so that an extra, slow pion
emitted from the first decay vertex can combine with the
$4$-momenta of the second vertex particles to form a $D^*$.
Finally all particles from the two decay vertices should
reconstruct a narrow $T(\tq)$ state below $2\,m_{D}$.

(b) $ m_{D^*} + m_{D} \geq \, m_{T(\tq)} \, >\, 2\,m_{D}$ or
$0 > \e\,\geq \, -\, 140 \, MeV$ \\
In this case we will have an electromagnetic decay:
\begin{equation}
T(\tq)^+ \to D^+ + D^0 + \gamma \,, \label{rd}
\end{equation}
with two weak decay vertices due to $D^+$ and $D^0$ decays. While
the state will be narrow,  reconstructing the invariant mass,
$m_{D^+ \, D^0 \, \gamma}$, and looking for a sharp peak requires
identifying a relatively soft photon ($E_{\gamma} \leq 140 \,
MeV$ in the $T(\tq)$ rest frame) emerging from the primary decay
vertex.

(c) $m_{T(\tq)} >  m_{D^*} + m_{D} $ or $\e \geq 0$ \\
Here we clearly have $T(\tq) \to D^{*0} + D^+$ (or
$D^{*+} + D^0$). The $D^{*}$ will decay into $D + \pi$ at the
primary vertex and the two $D$'s will next decay weakly at
separate vertices. Thus we attempt to reconstruct the two $D$'s
and the $D^*$ and finally look for an overall peak in the
$D^*\,D$ invariant mass distribution. This last constraint may
not be very helpful (and the $T(\tq)$ can be altogether missed)
if its width $\Gamma (T(\tq))$ (substantially) exceeds the experimental
resolution which we optimistically take to be ${\cal O}(10\, MeV)$.

Since $T(\tq, 1^{+}) \to D^*\, D$ is an $S$-wave decay and no new
quark pairs need to be created (as in $K^* \to K\, \pi$, for instance) one
might expect a large decay width, $\Gamma \approx 300 \, MeV$, as
is the case with $\bar{q}_i\,q_j\, \bar{q}_l\,q_k$ exotics made of
light quarks. Two factors may, however, reduce $\Gamma(T(\tq))$.
First we have a two-body decay phase space which is proportional
to $\beta$ or $\beta^*$, the velocity of $D$ or $D^*$ in the
$T(\tq)$'s rest frame, which unlike in the decays of light
exotics may be significantly less than one:
\begin{equation}
\beta_{D} \approx \beta_{D^*} \approx \sqrt{\e\over m_{D}}
\approx 0.13 \, \left({\e \over 30 \, MeV}\right)^{1/2} \, .
\label{beta}
\end{equation}

Second, it may well be that the physical $1^{+}$ hadron of
interest has a relatively small deuson component, $|\alpha|^2$:
\begin{equation}
| \Psi_{\tq} \, (physical)\,\rangle = \sqrt{1-|\alpha|^2}\, |
T(\tq) \rangle + \alpha\, |(D^*\,D)\, deuson \,\rangle \,,
\label{psi}
\end{equation}
with $T(\tq)$ being the ``one bag'' genuine four quark state.
Since only the deuson component readily falls apart into $D^* +
D$ we may then have a further suppression by a factor of
$|\alpha|^2$. If  $|\alpha|^2 \leq {1\over 3} - {1\over 4}$ the
joint $|\alpha|^2 \, \beta$ effect reduces the decay rate from 
$\Gamma \approx 300 \, MeV$ to $\Gamma \approx 13-19 \, MeV$.

Note, since the state of interest is an $S$-wave of the $D + D^*$ we may lack
any repulsive interaction---akin to Coulomb repulsion in fission or 
$\alpha$ decays---to generate a resonance in the first place \cite{JB}.
Rather we may
have, as in the $I=1$ $S$-wave nucleon-nucleon scattering, a ``virtual bound 
state''. The experimental manifestation of the latter---a strong enhancement 
at the $D\,D^*$ threshold, may be still sufficiently striking.
The above lifetime estimates would then still pertain to the width of this
enhancement.

In principle we can also have $c\,\bar{u}\,\bar{c}\,d$ tetraquark
(and/or meson-antimeson, $D^* \, \bar{D}$, $D\, \bar{D}^*$
deusons) with ``hidden'' charm. Such combinations will be much
more easier to produce as only one $\bar{c}\, c$ pair needs to be
created. Naively one would expect such states to decay very
quickly into $J/\Psi + \pi^+$. T$\ddot{\rm{o}}$rnqvist, whose main concern
was actually meson-antimeson binding did, however, suggest that a
primary extended deuson $D^*\, \bar{D}$ state would have
little overlap with the $J/\Psi$ (compact $\bar{c}\,c$) state.
This would then make for a relatively narrow $D^*\, \bar{D}$ state
which still could be nicely identified via its unique ($J/\Psi +
\pi^+$) decay mode at a specific $J/\Psi + \pi^+$ invariant mass.

The $\ccu/\ccd$ double charmed baryons have presumably been seen
at FNAL \cite{SELEX}. The $T(\tq)$ is most likely
produced there with high lab momenta. The proximity of the
$T(\tq)$ mass to the $D^*\,D$ threshold cause the two subsequent
weak decay vertices to be very precisely aligned with the initial
interaction vertex---a feature which will be most helpful in a
tetraquark hunt.

The much lower energy $e^+ \, e^-$ colliders can also serve as
promising search grounds with a much cleaner environment. Thus at
$\sqrt{s} \approx 10.7 \, GeV$ ($\approx m_{\Upsilon(4s)}$) the
primary virtual photon interaction yields in about $25 \%$ of
all cases a $\bar{c}\,c$ pair. This then will vastly increase the
number of events in which we have two charmed quark pairs to be
${\cal O}(1-0.25)$, {\it i.e.} ${\cal O}(10^6)$ events \cite{prv}.
The events in which $(J/\Psi)$ will not form will have typical
$D^{(*)}\, \bar{D}^{(*)}\, D^{(*)}\, \bar{D}^{(*)}$ final states
($D^{(*)}$ indicates $D$ or $D^*$) with a few extra particles due
to the limited phase space ($4\, m_{D^*} \approx 8.4 \, GeV$!).
The systematic search of peaks in $D^* \, D^{(*)}$ or
$\bar{D}^{(*)} \, \bar{D}$ invariant mass distribution could thus
be feasible despite the large combinatorial background.

Peaks in $D^* \, \bar{D}^{(*)}$ mass distribution could also
occur. These should manifest in the much cleaner ($J/\Psi +
\pi$) channel. Hopefully by $D^*$ (or $D^{**}$) cascade decays and by $K^+$,
$K^-$ separation one will be able to distinguish
$D^{(*)}\, D^{(*)}$ from $D^{(*)}\, \bar{D}^{(*)}$ pairs.

Finally we note that the Brown-Hanbury-Twiss effect favoring
$D^0\, D^0$ (and $D^+ \, D^+$) pairs with small relative momenta
should not be operative here as the two pseudoscalar $D$'s do not
emerge from the same vertex (even the $D$'s from the $D^* \to
D\,\pi$ decay would emerge several hundred Fermies away from the
primary vertex because of the narrow $D^*$'s width).

\section{Summary and Conclusions}

We have presented various estimates pertaining to a
possible tetraquark $T(\tq) (1^{+})$ state. If the SELEX second
peak corresponds to the lightest doubly charmed baryon,
then our estimates of the mass of $T(\tq) (1^{+})$ are close to
the $D\,D^*$ threshold. We have also discussed signatures for
various possible masses.

\section*{Acknowledgments}
B.G. was supported by the U.S. Department of Energy, under
Grants No. DE-FG02-93ER-40762 and No. DE-FG03-01ER-41109. S.N. acknowledges a
grant of the Israeli Academy
of Science. One of the authors, S.N., would like to thank M.A. Moinester and
H.J. Lipkin for many discussions. Both authors would like to acknowledge most
helpful discussions with M.K. Banerjee, T.D. Cohen and S.J. Wallace which, in
particular, helped to clarify some misconceptions regarding the OPEP.

\end{document}